# Microwave Photonic Exciter Unit for Radar System

Vishal Maheshwari, K.Sreenivasulu, Mohit Kumar, Kishan Sharma
Electronics and Radar Development Establishment (LRDE)
C.V. Raman Nagar, Bangalore - 560 093, INDIA
vishalmaheshwari1990@gmail.com

*Abstract*

Modern radar systems requires low phase noise and long term phase stable radar carrier signals for high resolution imaging, micro Doppler signatures applications. High stable signal generation through conventional crystal followed by frequency multiplier techniques are limited phase noise performance under vibration conditions. Optoelectronic oscillators (OEOs) offer low phase noise and long-term phase stability compared to the conventional oscillators. In this design of Radar Exciter unit based on microwave photonics components viz., OEOs, frequency dividers, optical filters, optical delay lines, optical arbitrary waveform generators are discussed. The microwave photonics offers frequency independent Exciter design with tunable filters and tunable laser sources. In this paper the performance comparison of microwave photonic-based exciter unit also discussed.

## I INTRODUCTION

As the technologies are progressing, It has become mandatory for all to built fast, distortion less, secure, reliable and light weight communicative radar system. Hence optical field fulfills such requirement up to large extant due to its high bandwidth, less losses, no EMI/EMC problem etc.

Mechanical, Electromagnetic and Atomic Oscillator cannot be used in such application where very high stability and spectral purity are required. So this paper introduce about Optoelectronic Oscillator (OEO) for generating stable and spectrally pure periodic signal along with optical filter, optical arbitrary waveform generator, frequency divider. OEO was invented in 1994 by Yao and Maleki which converts light energy from a continuous laser source to a periodically varying sine/cosine microwave signal with particularly low phase noise and very high quality factor. It consists of Continuous Wave pump laser of 1550nm and feedback circuit including optical modulator, fiber, photo detector and narrow band pass filter. A comprehensive simulation model is developed using optical design software OptSim 5.3 (RSoft Module) as shown in fig 1. Optical arbitrary wave generator is used in exciter to generate optical waveform of desire frequency and shape, which is up converted to high frequency signal. In this process optical filter is also required along with optical fiber in order to select optical signal of particular frequency.

Microwave photonic exciter unit for radar system is implemented using optical arbitrary waveform generator, optical modulator, optical filter and OEO as shown in figure 1(b) which is equivalent to conventional exciter unit as shown in figure 1(a).

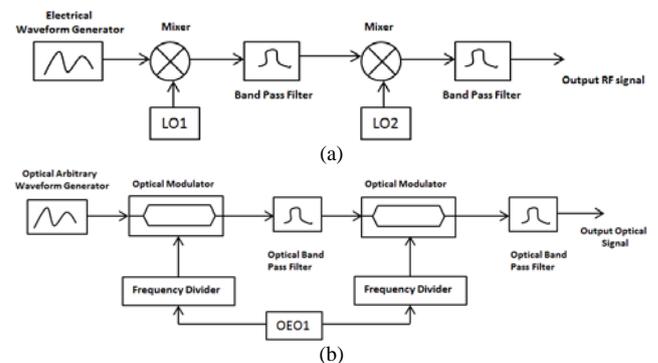

Figure 1: Exciter chain in radar (a) conventional using electrical component [1] (b) using optical interface

In this paper we first describe the basic principle of OEO along with theoretical overview of optical arbitrary waveform generator, optical filter and gives results achieved during simulation for OEO on OptSim v5.3 platform. Further this paper will be concluded with comparison of optical based exciter unit with conventional RF exciter.

## II BASIC PTINCIPLES

*1. OPTOELECTIC OSCILLATOR (OEO)*

OEO is characterized in optical domain by having high quality factor and stable microwave electric signal in form of sine/cosine when continuous light wave is given to OEO system. OEO consists of pump laser, Machzender (MZ) Modulator, Fiber, Photo detector, Amplifier and Band pass filter as shown in figure 2.

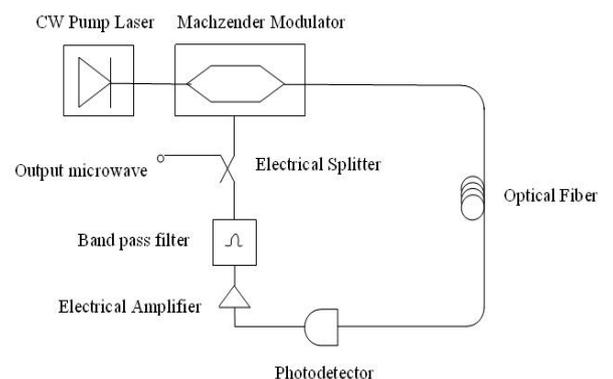

Figure 2: Schematic of Opto-Electronic Oscillator (OEO)

Its photonic components characterized with better efficiency, high speed, high bandwidth and low dispersion in the microwave frequency regime. This works on principle of converting light wave to electrical signal and optical fiber delay line works as energy storing element in system like electronic oscillator has Inductor and capacitor. We have used here 1550nm light because of less attenuation and less dispersion during communication. This 1550 nm light is introduced from continuous pump laser to MZ modulator which has its own bias and one more electrical signal. Intensity modulated light from modulator is passed through fiber which is acting as energy storing element depending on length of fiber. Output from optical fiber is passed through photo detector which converts all delayed optical signal to electrical signal. But photo detector generates harmonics during conversion process hence a narrow band pass filter is put to extract desire signal with high quality factor. And after amplification electrical signal is fed back to modulator and feedback loop is completed. This configuration supports self sustained oscillation at frequency determined by the fiber delay length and band pass filter property. Quality factor for such system can be calculated as $Q = 2\pi f \tau_d$, where f is frequency of electrical signal and $\tau_d$ is time delay occur due to fiber. Thus we have developed regenerative feedback approach to produce electrical stable sine signal[2]

Oscillation frequency is limited only by the characteristic of frequency response of the modulator and filter deign, which eliminates all other sustainable oscillation. One condition is needed for such oscillation, that adequate light input power is required. So for satisfying such condition we have put electric amplifier at feedback loop along with high power laser. The oscillator consists of an amplifier of gain G and a feedback transfer function β(f) in a closed loop. The gain G compensates for the losses, while β(f) selects the oscillation frequency. Barkhausen condition gives G.β(f) = 1.

*2. OPTICAL ARBITRARY WAVEFORM GENERATOR*

This is used to generate optical waveform of different frequency. This is made by using a Spatial Light Modulator (SLM) array to shape the spectrum of the broadband pulse. The approach is a coherent Fourier transform process where a temporal waveform was synthesized through manual control of optical phase.

A broadband optical source is produced by amplifying the output of a modelocked laser and passing it through a SuperContinuum (SC) fiber. Optical nonlinearities in the SC fiber cause broadening of the optical spectrum. Next, a spatial light modulator filters and shapes the spectra according to the desired optical waveform. We use a 4-f grating and lens apparatus such that each wavelength will be focused and incident normal onto the SLM plane.[3]

*3. OPTICAL FILTER*

Optical filter is a device to selectively transmit light in particular range of frequencies, while absorbing or reflecting back unwanted frequency band. That is basically called as interference filter also, made by coating substrate with a series of optical coating. Their layers form a sequential series of reflective cavities that resonate with the desired wavelengths. Other wavelengths destructively cancel or reflect as the peaks and troughs of the waves overlap. Its property is dependent on substrate, thickness and sequence of coating. Optical filter is also classified like electrical filter. Optical filters are named as Short Pass Filter: Transmit light of frequency below cut-off frequency; Long Pass Filter: Transmit light of frequency above cut-off frequency; Band Pass Filter: Transmit light having certain range of frequency. These are similar to electrical Low Pass, High Pass and Band Pass filter respectively.[4]

### III SIMULATION OF OEO

*1. OPTSIM v5.3 (OPTICAL SIMULATOR)*

OptSim, RSoft optical design suite, is for optical simulation; designed for optical communication systems and simulates them to determine their performance given various component parameters which works on both Windows and UNIX platforms. It includes the most advanced component models and simulation algorithms, validated and used for research documented in numerous peer reviewed professional publications, to guarantee the highest possible accuracy and real-world results. OptSim represents an optical communication system as an interconnected set of blocks, with each block representing a component or subsystem in the communication system. This allows users to design and simulate optical interconnects with electrical system at signal level propagation. Each block is simulated independently using the parameters specified by the user for that block and the signal information passed into it from other blocks. This is known as a block-oriented simulation methodology.[5]

*2. SIMULATION MODAL DESIGN ON OPTSIM*

We have designed OEO in two layers as shown in figure 3. Figure 3(a) shows upper layer of whole schematic and figure 3(b) is made to provide feedback as shown in figure 2. Under this feedback loop, electrical input is given to modulator and which will be converted to our desire electrical signal after some iteration. But for starting the simulation we have given a random electrical signal to modulator which we generated from optical photo detector on passing optical random signal. Basically this feedback is kind of infinite loop but for the purpose of simulation we have fixed it to six number of iteration.

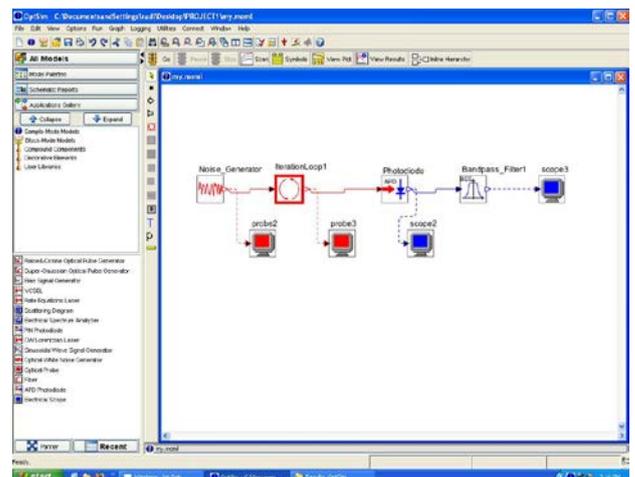

(a)

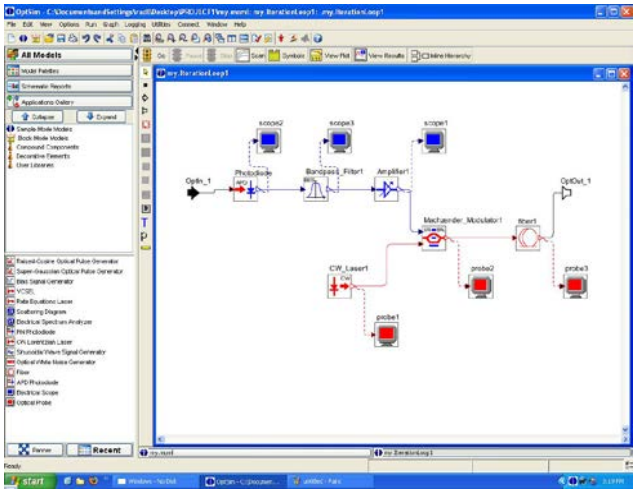
(b)

Figure 3. (a) Upper layer of Design Schematic of OEO (b) Feedback Design built under loop block shown in 1st layer

1550 nm laser of 3dBm power is given to modulator along with electrical input. Modulator's output goes to 10 km long fiber (L = 10 KM) whose output connected to photo detector having 0.9 A/W responsivity. Electrical output from detector is then filtered out at 10GHz frequency using very narrow and high quality factor band pass filter whose output is given to modulator after amplification. This simulation is made for 6 feedback cycle/loop only which can be customized by user to more number of cycle. Electrical and optical probe are shown in above schematic, is used to see electrical and optical signal. All red connections are optical connection and blue parts are related to electrical connection.

*3. RESULTS*

Following are results obtained on simulation of above modal. Red and blue signal are representing optical signal and electrical signal respectively.

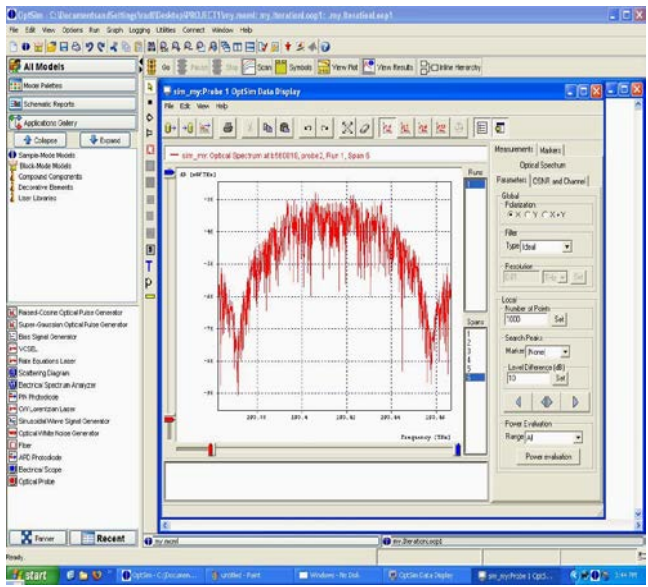

Figure 4. MZ Modulator output after 6th loop

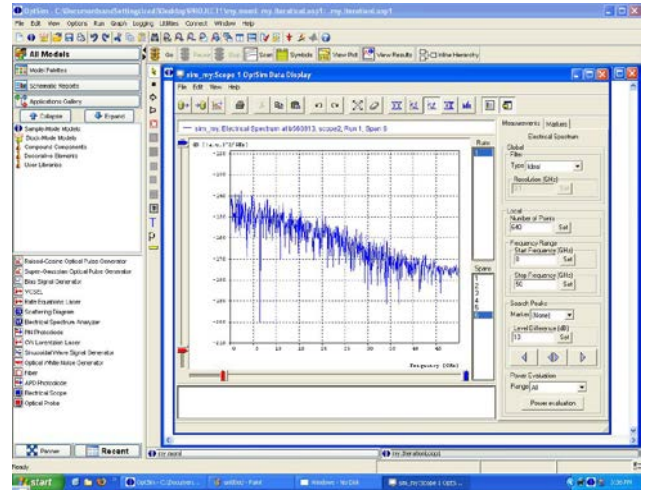

Figure 5. Photo Detector output after 6th loop

There are so many peaks are coming into picture as shown in figure 5. These peaks are representing harmonics generated by photo detector during conversion of light wave to electrical signal.

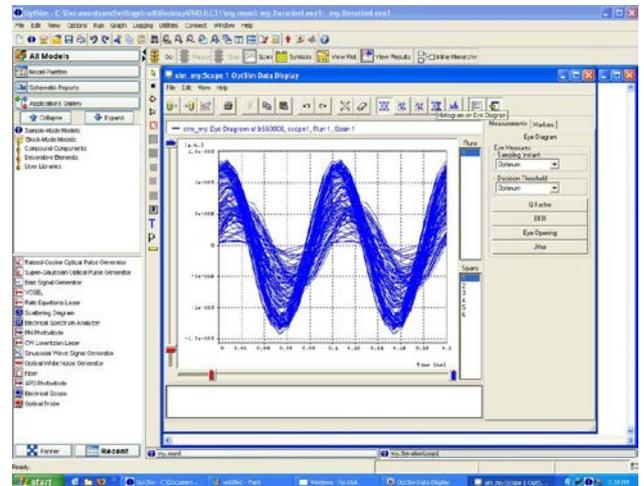
(a)

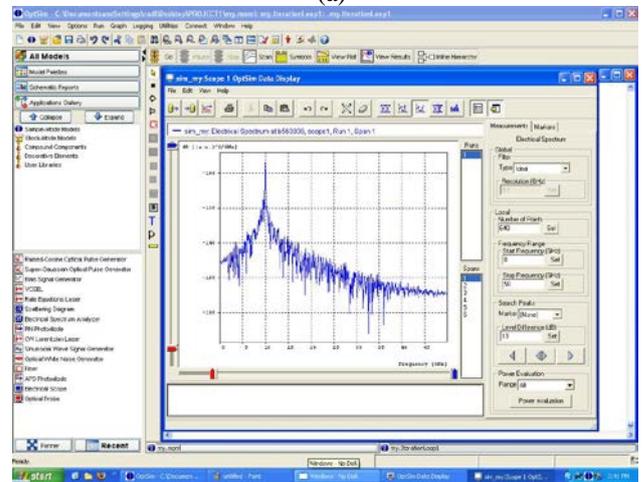
(b)

Figure 6. Output from OEO system after 1st loop only (a) in time domain (b) in frequency domain

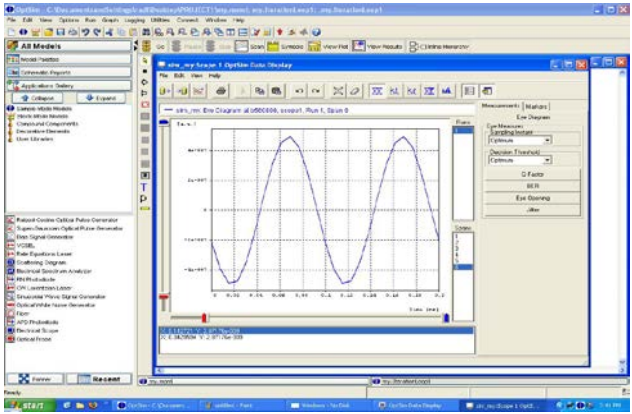

(a)

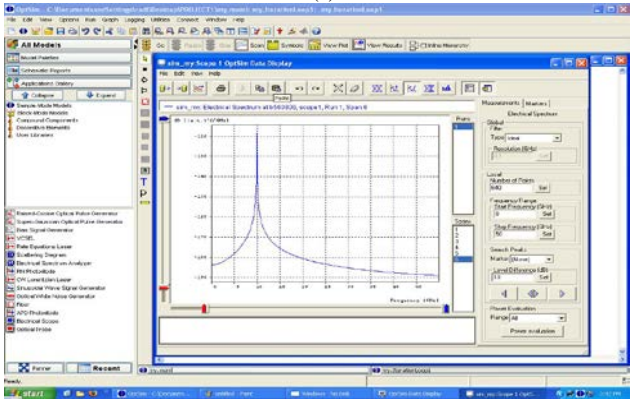

(b)

Figure 7. Output from OEO system after 6th loop (a) in time domain (b) in frequency domain

Thus it can be compared between figure 6 and figure 7 that initially output was not spectrally pure and stable. And so after 6$^{th}$ iteration, output is perfectly sine wave with frequency of 10 GHz (microwave region) as shown in figure 7, which is stable and having very low phase noise as compare to figure 6. Its Q factor is of the order of $10^6$ which can be calculated also as $Q = 2\pi f \tau_d$. Here f = 10GHz and $\tau_d$ = L/C (L is length of fiber = 10KM and C is speed of light = $3 \times 10^8$ m/s). So calculated value of Q factor is $2 \times 10^6$, which is equivalent to simulated result.

Simulation for Optical arbitrary waveform generator and Optical can also be done in same manner. Theoretical study is carried out on these optical components which provide a good picture for their implementation.

## IV CONCLUSION

We have introduced a highly stable, highly spectral purity optoelectronic oscillator to achieve low phase noise electrical microwave signal. We have generated 10GHz sine signal with high quality factor and good phase noise using OEO feedback system. Output of OEO becomes stable after 6 iterations. OEO simulated using blocks of CW laser, MZ Modulator, fiber delay line, photo detector, narrow band pass filter and amplifier on RSoft product, OptSim v5.3 platform. Paper gives brief introduction of Optical arbitrary waveform generator and Optical filter and their use in photonic exciter unit of radar. This concludes that conventional exciter unit can be replaced with optical exciter unit in equivalent manner. This provides weight reduction of system, spectral purity in waveform because of high Q factor (~$10^6$) which we obtained using very narrow band pass filter, fast communication channel with high bandwidth with no EMI/EMC interferences, Less losses in compare to conventional system because optical cable losses is of the order of 0.2-0.3 dB/Km and whereas electrical cables losses is of the order of 1-2 dB/m. Good phase noise is achieved. Phase noise does not degrade under vibration condition in optical system but it degrades to 0.04g$^2$/Hz in conventional system.

## BIO DATA OF AUTHOR(S)

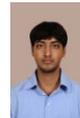
**Vishal Maheshwari** born on 18$^{th}$ December 1990 obtained his B.Tech degree in Electrical Engineering stream from IIT Gandhinagar in 2012. He is currently working as scientist at Electronics and Radar Development Establishment (LRDE), Bengaluru. Area of specialization is in Digital communication and optical realization

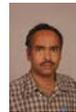
**K Sreenivasulu** received his Diploma in Electronics and Communication Engineering from S.V. Government Polytechnic, Tirupati, Andhra Pradesh in the year 1987. He received his B.Tech degree in Electronics and Communication Engineering from Jawaharlal Nehru Technological University, Hyderabad in the year 1995. He received M.E. degree in Micro Electronics Systems from Indian Institute Of Science, Bangalore in 2004. He started his professional career as Electronic Assistant in Civil Aviation Department where he worked from 1990 to 1995. Since 1996 he has been working as Scientist in Electronics and Radar Development Establishment (LRDE), Bangalore. His area of work has been design and development of RF and Microwave sub-systems, Digital Radar system, Beam Steering Controller for Active Aperture Array Radars. His interests include VLSI Systems and Programmable Controllers.

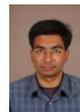
**Mohit Kumar** born on 7$^{th}$ October 1980 obtained B.Tech degree in Electronics and Communication from NIT,Jalandhar in 2002. He has completed his MTech from IIT Delhi in Communication Engg. in 2010. He is currently working as scientist at Electronics and Radar Development Establishment (LRDE), Bengaluru. Area of specialization is in Digital Radar Receiver design.

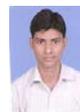
**Kishan Sharma** born on 5th September 1985 obtained BE degree in Electronics from Jiwaji University Gwalior in 2008. He is currently working as scientist at Electronics and Radar Development Establishment (LRDE), Bengaluru. Area of specialization is in RF Receivers for Radars.